\documentstyle[epsf,12pt]{article}
\begin{document}
\title{ISOLATED STATES \thanks{Presented as a talk at XVII Symposium
on Nuclear Physics, Oaxtepec, Mexico, January 4-7, 1994}}
\author{ \underline{A.M.Shirokov} \\
Institute of Nuclear Physics, Moscow State University, \\
Moscow 119899, Russia \\
\and
Yu.F.Smirnov \\
Instituto de F\'{\i}sica, Universidad Nacional Aut\'{o}noma \\
de M\'{e}xico, Apdo. Postal 20-364, M\'{e}xico 20, D.F. \\
 and \\
 Institute of Nuclear Physics, Moscow State University,  \\
 Moscow 119899, Russia \\
\and
S.A.Zaytsev \\
Institute of Nuclear Physics, Moscow State University, \\
Moscow  119899, Russia }
\date{ }
\maketitle
\begin{abstract}
We show that a quantum system with nonlocal interaction can have bound
states of unusual type (isolated states (IS)). IS is a bound state that do
not generate a $S$-matrix pole. IS can have positive as well as negative
energy and can be treated as a generalization of bound states embedded in
continuum on the case of discrete spectrum states. Formation of IS in the
spectrum of quantum system is studied using a simple rank--2 separable
potential with harmonic oscillator formfactors. Some physical applications
are discussed, in particular, we propose separable $NN$ potential that
describes not only most important two-nucleon data (deuteron binding energy
and $s$-wave triplet and singlet scattering phases) but also the trinucleon
binding energy without making use of three-body forces.
\end{abstract}
\section{Introduction}
\par
     Phenomenological local potentials are conventionally used  in
nuclear physics for the description of $NN$ interaction, nucleon-cluster
or cluster-cluster interactions,  etc.  Nevertheless,  interaction
between quantum particles is nonlocal in nature.  The  nonlocality
arises from the  composite  structure  of  interacting  particles,
e.g., from the quark structure of nucleons; it appears as a result
of relativistic effects or of the truncation of a model space, etc.
\par
    The nonlocality can give rise to some peculiarities in scattering
and/or in the structure of bound states. We  shall  show  that  a
system with a nonlocal interaction can  have  a  bound  state  of  a
special type that we shall refer to as isolated state (IS). IS  is  a  state
with dying asymptotics like a usual bound state but IS is  not  in
correspondence with any of $S$-matrix poles. IS can have negative or
positive energy; in the later case, IS appears to be a bound  state
embedded in continuum (BSEC) that have been studied both for local \cite{9}
and nonlocal \cite{10} interactions.
\par
     We shall present a model  finite-rank  separable  interaction
giving  rise  to  IS.  Exact  solutions   of   the   corresponding
Schr\"{o}dinger equation can be easily found in the $J$-matrix formulation
of the scattering theory \cite{1}--\cite{4}. The $J$-matrix approach is based
on the Fourier-expansion of continuum  spectra  wave  functions  in
terms of $L^2$  basis functions. In particular, the harmonic  oscillator
eigenfunctions as well as the so-called  Sturm-Laguerre  functions
can be used as the basis functions in the method. The important  point
of this approach is the fact that the matrix of the kinetic energy
operator $\hat{T}$ is of a tridiagonal form on the oscillator or
Sturm-Laguerre basis, i.e., the kinetic energy matrix is a Jacobi
matrix\footnote{Jacobi matrix is by definition a tridiagonal Hermitean
matrix.}
(or, shortly, $J$-matrix). The main advantage  of  this  formalism  in
comparison with currently used formalisms in the scattering  theory  is
the fact that it deals with algebraic rather than differential  or
integrodifferential equations.
\section{$J$-matrix formalism}
     The radial wave  function  $\Psi^{l}_{E}(r)$  satisfies  the
Schr\"{o}dinger equation,
\begin{equation}
(\hat{T} + \hat{V} - E)\Psi^{l}_{E}(r)=0,          \label{(1)}
\end{equation}
where $E$ is the energy, $l$ is the angular momentum, $\hat{T}$ is
the kinetic energy operator, while an interaction in the system  we
describe by a separable nonlocal potential,
\begin{equation}
\hat{V} = \sum_{n,n'=0}^{N} V_{nn'} \mid
\varphi_{nl}(r)\rangle\langle\varphi_{n'l}(r')\mid ,           \label{(2)}
\end{equation}
of the rank $N+1$ with the harmonic oscillator formfactors,
\begin{equation}
\varphi_{nl}(r) = (-1)^{n} \left[ \frac{2n!}{r_{0}\Gamma (n+l+\frac{3}{2})}
\right] ^{\frac{1}{2}} \left( \frac{r}{r_{0}} \right) ^{l+1} \exp \left(
- \frac{r^2}{2r^{2}_{0}} \right) L_{n}^{l+\frac{1}{2}} \left(
\frac{r^2}{r_{0}^{2}} \right).                          \label{(3)}
\end{equation}
Here $r_{0} =(\hbar /m\omega )$ is the oscillator radius, and
$L^{\alpha}_{n} (x)$ is a Laguerre polynomial. In the $J$-matrix method,
the wave function has a form of Fourier series in terms of $L^{2}$
functions (\ref{(3)}), i.e.,
\begin{equation}
\Psi ^{l}_{E} (r) = \sum_{n=0}^{\infty} X_{n} (E) \varphi_{nl}(r).
                 \label{(4)}
\end{equation}
\par
      The coefficients $X_{n} (E)$ of the expansion (\ref{(4)}) for $n\geq N$
are given by the formula,
\begin{equation}
X_{n} (E) = S_{nl}(p) \cos \delta _{l} + C_{nl}(p) \sin \delta_{l}
\; ,                                               \label{(5)}
\end{equation}
where $p=\sqrt{2E/\hbar\omega}$  is a momentum, and $S_{nl}(p),\; C_{nl}(p) $
are eigenvectors of the infinite tridiagonal matrix of the kinetic  energy
$T_{nn'}$. For calculation of $S_{nl}(p)$  and  $C_{nl}(p)$  the following
analytical expressions \cite{2,4} can be used,
\begin{equation}
S_{nl}(p) = \left[ \frac{2\Gamma (n+l+\frac{3}{2})}{\Gamma (n+1) } \right]
^{\frac{1}{2} } \frac{p^{l+1} }{\Gamma (l+\frac{3}{2}) } \exp
(-\frac{p^2}{2}) \: _{1}F_{1}(-n,\, l+\frac{3}{2};\, p^{2})
\: ,                                              \label{(6a)} \\
\end{equation}
\begin{eqnarray}
C_{nl}(p) = \left[ \frac{2\Gamma (n+1)}{\Gamma (n+l+\frac{3}{2}) }
\right] ^{\frac{1}{2} } \frac{(-1)^{l} }{p^{l} \Gamma (-l+\frac{1}{2}) }
\exp (-\frac{p^2}{2}) \nonumber \\
 \times _{1}F_{1}(-n-l-\frac{1}{2},\, -l+\frac{1}{2};\, p^{2})\:
.                                                \label{(6b)}
\end{eqnarray}
\par
     The coefficients $X_{n}(E)$ for $n<N$ may be found by the formula,
\begin{equation}
X_{n}(E) = \wp _{nN}(E)X_{N+1}(E)\: ,                     \label{(7)}
\end{equation}
where
\begin{equation}
\wp _{nn'}(E) = - \sum_{\mu } \frac{U^{\mu }_{n} U^{\mu }_{n'} }
{\varepsilon _{\mu} - E } T^{l}_{n',n'+1} \; ,               \label{(8)}
\end{equation}
$U$ is the transformation that diagonalizes the truncated Hamiltonian
matrix, $H^{N}_{nn'} =T^{l}_{nn'} + V^{l}_{nn'} \; (n,n'=0,1,...,N)$,
$U^{\mu}_{n}\; (n = 0,1,...,N)$ are eigenvectors and $\varepsilon _{\mu}$
are the corresponding eigenvalues of this matrix, i.e.,
\begin{equation}
[U^{+} H U]_{\mu\mu '} = \delta _{\mu\mu '}\varepsilon _{\mu}\;.
                     \label{U}
\end{equation}
\par
Setting $n=N$ in eq.(\ref{(7)}) and using expression (\ref{(5)})
both for $X_{N}(E)$ and $X_{N+1}(E)$, we obtain the following
formula for the phase shift $\delta _l$  in the
partial wave characterized by the angular momentum $l$,
\begin{equation}
\tan \delta _{l} = - \frac{S_{Nl}(p) - \wp _{NN}(E)S_{N+1,l}(p)}
{C_{Nl}(p) - \wp _{NN}(E)C_{N+1,l}(p)}\; .          \label{(9)}
\end{equation}
\par
     All the above expressions may  be  generalized  on  the  coupled
channel case \cite{2}, on the case of true few-body scattering
($A\to A$) \cite{5}, etc. The $J$-matrix approach allows one to analyze the
wave functions and other scattering characteristics in a simple and visual
manner. We shall demonstrate this feature with a simple model of IS that will
be proposed in the next section.
\section{Simple model of IS}
     Let us consider the case when the Hamiltonian matrix $H_{nn'}$ is of a
block-diagonal form in the harmonic oscillator basis, i.e., $H=H^{(1)}\oplus
H^{(2)}$, where the submatrix $H^{(1)}$ acts in a finite-dimensional subspace
of the basis functions (\ref{(3)}) with $n<M$. The infinite-dimensional
submatrix $H^{(2)}$
acts in an orthogonal supplement to this subspace. Bearing in mind
the finite-rank structure of the potential (\ref{(2)}), we  conclude  that
the Hamiltonian matrix $H_{nn'}$ is of the form presented  on  Fig. 1.
\begin{figure}[bht]
\unitlength=1.00mm
\special{em:linewidth 0.4pt}
\linethickness{0.4pt}
\begin{picture}(72.34,74.33)
\put(22.67,29.33){\framebox(29.67,30.00)[cc]{$H^{(2)}$}}
\put(7.67,59.33){\framebox(15.00,15.00)[cc]{$H^{(1)}$}}
\put(52.34,29.33){\line(1,-1){20.00}}
\put(48.34,29.33){\line(1,-1){20.33}}
\put(52.34,33.33){\line(1,-1){20.00}}
\put(14.01,46.66){\makebox(0,0)[cc]{{\large 0}}}
\put(37.34,67.33){\makebox(0,0)[cc]{{\large 0}}}
\put(37.34,17.33){\makebox(0,0)[cc]{{\large 0}}}
\put(66.67,42.66){\makebox(0,0)[cc]{{\large 0}}}
\end{picture}
\caption{The structure of the Hamiltonian matrix.}
\end{figure}
An infinite tridiagonal tail of the Hamiltonian matrix represents
the kinetic energy matrix  elements, the remaining non-zero matrix elements
are displayed graphically as the boxes representing submatrices $H^{(1)}$
and $H^{(2)}$ (we include  the  infinite  tridiagonal  kinetic
energy tail in the submatrix $H^{(2)}$).
\par
Obviously, the eigenvectors of
the submatrix $H^{(1)}$ are eigenvectors of the total Hamiltonian matrix $H$,
too. The corresponding wave functions are dying as $r \to \infty$
similarly to the wave functions of bound states since
they are linear combinations of a finite number of the functions (\ref{(3)}).
Nevertheless, the eigenstates of the submatrix $H^{(1)}$ are characterized
by wave functions with unusual asymptotics $\sim$$
e^{-\frac{r^{2}}{2r_{0}}}$ instead of  standard ones $\sim$$e^{-\alpha r}$.
Submatrix $H^{(1)}$ may have positive or/and negative eigenvalues
$\varepsilon ^{(1)}_{\mu}$. If $\varepsilon ^{(1)}_{\mu} > 0$, the
corresponding state appears to be BSEC. If $\varepsilon ^{(1)}_{\mu} < 0$,
the corresponding state appears to be a bound state of a special type.
All eigenstates of the submatrix $H^{(1)}$ will  be  referred
to as isolated states (IS), because they are isolated
completely from the scattering states.
\par
The continuum spectrum states with oscillation--type asymptotics of the wave
function as well as the usual bound states with $e^{-\alpha r}$--type
asymptotics of the wave functions, are expressed as infinite series
(\ref{(4)}). They are, obliviously, eigenstates of the submatrix $H^{(2)}$.
The eigenvectors of the submatrix $H^{(2)}$ being the
wave functions of scattering and usual bound states in the harmonic
oscillator representation are, obviously, orthogonal to the eigenvectors
of the submatrix $H^{(1)}$. Thus, the scattering and usual bound state
wave functions have node(s)  at  a
small distance ($\sim$$r_0$) in the presence of IS(es) and are similar
to the wave  function for the potential with forbidden states \cite{6}.
\par
The asymptotic behavior of scattering state wave functions is characterized
completely by the $S$-matrix \cite{8}. So, the structure of the
$S$-matrix is dictated by
the infinite-dimensional submatrix $H^{(2)}$ and does not depend on the
submatrix $H^{(1)}$. The IS energies $\varepsilon ^{(1)}_{\mu}$, on the other
hand, are dictated by the submatrix $H^{(1)}$ and do not depend on the
submatrix $H^{(2)}$. Varying matrix elements $H^{(1)}_{nn'}$ of the
submatrix $H^{(1)}$ one causes variation of the IS energies
$\varepsilon ^{(1)}_{\mu}$ without any variation of the $S$-matrix. Thus,
the energy of IS $\varepsilon ^{(1)}_{\mu}$ is not in correspondence with
the location of $S$-matrix poles. Using symmetry properties of $S$-matrix as
a function of complex momentum $p$ \cite{8}, it is easy to show \cite{Sm}
that the energy of BSEC is not in correspondence with any of $S$-matrix
poles. An interesting new point, so far as we know never discussed in
literature, is the appearance of the discrete spectrum
states, i.e., ISes with negative energy $\varepsilon ^{(1)}_{\mu}<0$, that
are divorced from $S$-matrix poles. So, IS being the state characterized by
the asymptotically dying wave function and the energy at which the $S$-matrix
has not a pole, can been treated as a generalization of BSEC on the case of
discrete spectrum states.
\par
    We shall examine in more detail the formation of IS in the spectrum of
the quantum system with nonlocal interaction using as an example a simple
analytically solvable model. The simplest realization of the situation
depicted on the Fig. 1, corresponds to the
case when $H^{(1)}$ is a 1$\times$1 matrix and the separable potential
(\ref{(2)})  is
of the rank 2, i.e., $N=1$. In this  case,  the  appearance  of  IS  is
connected with cancellation of the potential  energy  matrix  elements
$V_{01} = V_{10}$  with the kinetic energy  matrix  elements
$T_{01} = T_{10} = -V_{01}$,  i.e., $H_{01} = H_{10} = 0$. The energy
$\varepsilon _{0}$ of the IS  is  equal  to  the  diagonal  matrix
element $H_{00}$, $\varepsilon _{0} = H_{00}$. It should be mentioned,
that $H_{00}$ can  take
an arbitrary value in our model.  Using
eqs. (\ref{(5)}--\ref{(9)}) we obtain the following expression for the
phase shift $\delta _l$ :
\begin{displaymath}
\tan \delta _{l} =
 - \frac{S_{0l}(p) \left\{ \left[
V_{11}(\varepsilon _{0} - E) - \beta  \right] (T_{00} - E)
+ T^{2}_{01} (\varepsilon _{0} - E ) \right\} }{C_{0l}(p) \left\{ \left[
V_{11}(\varepsilon _{0} - E) - \beta  \right] (T_{00} - E)
+ T^{2}_{01} (\varepsilon _{0} - E ) \right\} -
\frac{p[V_{11}(\varepsilon _{0} -E) - \beta ]}{\pi S_{01}(p) } },
%   \label{(tan)}
\end{displaymath}
where $\beta \equiv H^{2}_{01}$.
%\begin{eqnarray*}
%{\begin{array}{l}
%\tan \delta _{l} =  \\
%\end{array} }  \\
%%\begin{array}{r}
% - \frac{S_{0l}(p) \left\{ \left[
%V_{11}(\varepsilon _{0} - E) - H^{2}_{01} \right] (T_{00} - E)
%+ T^{2}_{01} (\varepsilon _{0} - E ) \right\} }{C_{0l}(p) \left\{ \left[
%V_{11}(\varepsilon _{0} - E) - H^{2}_{01} \right] (T_{00} - E)
%+ T^{2}_{01} (\varepsilon _{0} - E ) \right\} -
%\frac{p[V_{11}(\varepsilon _{0} -E) - H^{2}_{01}]}{\pi S_{01}(p) } }
%\; .
%% \nonumber
%%      \label{(tan)}
%%\end{array}
%\end{eqnarray*}
\par
    Suppose  $\varepsilon _{0} >0$  and $V_{11} > 0$.
The  evolution  of  the  $s$-wave  phase  shift $\delta_{0}(p)$   as
$\beta = H_{01}^2 = H_{10}^2$ tends to zero is represented at the Fig. 2.
\begin{figure}[tb]
\epsfverbosetrue
\epsfysize=10cm
\epsfbox{bsecps.eps}
%\vspace{10cm}
\caption{ The evolution of the phase shift $\delta_{0}(p)$   as
$\beta = H_{01}^2 = H_{10}^2$ tends to zero. Dashed, dotted, and solid
curves correspond to the values of $\beta$ equal to $\beta_1$, $\beta_2$,
and $\beta_3$, respectively; $\beta_{1} > \beta_{2} > \beta_{3} = 0$.}
\end{figure}
If $\beta \neq 0$
there is a resonance at energy $E \approx \varepsilon_0$
of  the  width  $\Gamma$ that  becomes
smaller and smaller as $\beta$ tends to zero. At $\beta =0$
BSEC arises as  the
resonance of zero width producing the jump of the height $\pi$ of  the
phase shift $\delta_0$  curve at the point $E=\varepsilon_0$.
This spurious jump should be
omitted resulting in the $\delta_{0}(0)$  increase of extra $\pi$. Thus,
applying the Levinson theorem \cite{8} to the system pertaining IS, one
should treat it as a usual bound state. Such behavior of the
phase  shift  is  typical  for systems pertaining BSEC that have been
studied in various models \cite{8}--\cite{7}. Thus, our model represents an
alternative simple analytical approach in the study of BSEC. Such
behavior of $\delta_{l}(p)$ is also typical for the so-called
orthogonal  scattering or for the scattering
on the potential with forbidden state.
\par
     As for the S-matrix, it is given by the following expression:
\begin{equation}
S_{l} =
 - \frac{C^{(-)}_{0l}(p) \left\{ \left[
V_{11}(\varepsilon _{0} - E) - \beta  \right] (T_{00} - E)
+ T^{2}_{01} (\varepsilon _{0} - E ) \right\}
-\frac {p[V_{11}(\varepsilon _{0} -E) - \beta ]}{\pi S_{01}(p) }}
 {C^{(+)}_{0l}(p) \left\{ \left[
V_{11}(\varepsilon _{0} - E) - \beta  \right] (T_{00} - E)
+ T^{2}_{01} (\varepsilon _{0} - E ) \right\} -
\frac{p[V_{11}(\varepsilon _{0} -E) - \beta ]}{\pi S_{01}(p) }
},                            \label{(10)}
\end{equation}
where $C^{(\pm )}_{nl}(p)=C_{nl}(p) \pm iS_{nl}(p)$. The single
$S$-matrix pole in the non-physical sheet tends
to  the  real  energy  point  $E=\varepsilon_0$  as $\beta$ tends to  zero.
It is inherent to situation with BSEC.  However,  in  the
limit $\beta =0$, the factor $(\varepsilon_{0} -E)$
in the numerator of (\ref{(10)}) cancels the same
factor in the denominator and  the  singularity  in  the
point $E=\varepsilon_{0}$  in (\ref{(10)}) disappears,
i.e., the $S$-matrix in  the  presence
of BSEC has no poles. Nontrivial result is that if IS  belongs  to
the discrete spectrum (i.e., $\varepsilon_{0} <0$),
it  doesn't  generate   $S$-matrix pole, too.
\par
     The above results can be generalized as a following
\par
{\underline  Theorem.} Let all the eigenvalues of the truncated Hamiltonian
matrix $H^N$  be non-degenerate. Then  isolated  states  occur  in  the
spectrum of the system with rank--$(N+1)$ separable  interaction
(\ref{(2)})
if and only if the truncated matrix $H^N$   and  its  principal  minor
$H^{N-1}$  of the rank $(N-1)$ have common eigenvalues. The  number $\nu$ of
the coinciding eigenvalues is equal  to  the  number  of  isolated
states. These eigenvalues  and  corresponding  eigenfunctions  are
just the energies and the wave functions of the  isolated  states,
respectively.
\par
     It means that each IS is a linear combination of  the  finite
number of the harmonic oscillator functions $\varphi_{nl}(r)$
$( n \leq N-1 )$ and
its asymptotical behavior $\sim \exp (-\frac{r}{2r_{0}})$
is not connected with its energy.
\par
     The equivalent formulation of the theorem is:
\par
     The IS with the energy $\varepsilon_{\mu}$  exist if
and only  if  some  eigenvector $U^{\mu}$  of the matrix $H^N$
has the last component $U^{\mu}_{N} = 0$.
\section{Discussion}
\par
     The most interesting feature of the IS is the fact that  they
are not in correspondence with $S$-matrix poles.  This
impressive property  we use as a definition of IS:  IS is the
eigenstate of the Hamiltonian with dying asymptotics of the wave function
that do not  generate  $S$-matrix pole.
\par
     The results given  above  can  be strightforwardly generalized
on  the  case of coupled
channels problem, on the case of separable nonlocal potentials  of  finite
rank with the Sturm-Laguerre formfactors, etc.
\par
     As for physical applications, it should be noted that the  IS
appears in variational calculations using the harmonic  oscillator
basis. It is well-known \cite{DF,11} (the Dubovoy-Flores theorem)
that the lowest eigenvalue $\varepsilon_{0}^{N}(\omega )$  of
the matrix $H^N$  coincides with the lowest eigenvalue
$\varepsilon_{0}^{N}(\omega )$  of  the
matrix $H^{N-1}$  if the parameter $\hbar \omega$
of the basis  functions  minimizes
the eigenvalue $\varepsilon_{0}^{N-1}(\omega )$.
According to our theorem,  it  means  that
the corresponding eigenstate  $\Phi_0$  will  be  IS.
\par
 Another  physical
application may concern the $NN$ potential that should be
nonlocal from the point of view of the  quark  model.  A  simple  two-color
"toy" model of $NN$ interaction considered in Ref. \cite{12}  yields  the
separable nonlocal potential of  the  type  described  above  with
$V_{00}  >0$ and $V_{11}  =0$ while $V_{01}  =V_{10}$
has a  tendency  to  cancel
$T_{01}  =T_{10}$. Thus, the non-relativistic $NN$
scattering  problem  arising
in the quark resonating group model is close to the situation with
BSEC. If IS is really inherent to the $NN$-interaction,  it  may  be
used for  explanation  of  the  impressive  success  of  the  deep
attractive potential  with  forbidden  states  description  of $NN$
scattering \cite{6}.
     To illustrate the usefulness of using potentials with IS  for
the description of $NN$ interaction, we make fit  to  $NN$  scattering
data by the potential of the type (\ref{(2)}).  The  singlet  and  triplet
$s$-wave $NN$ scattering phases are  well  reproduced  by  second-rank
($N=1$) potential (\ref{(2)}) with  $V^{s}_{11} =-365.75$  MeV  and
$V^{t}_{11}  =-407.56$  MeV,
$\hbar \omega =500$ MeV;  $V^{s}_{10}  =V^{s}_{01}$   and
$V^{t}_{10}  =V^{t}_{01}$   are  fixed  by  the  condition
 $V^{s}_{10}  =V^{s}_{01} =V^{t}_{10}  =V^{t}_{01} =-T_{01}$.
Deuteron binding energy $E_d$  and deuteron rms
radius $<r^{2}_{0}>^{1/2}$  are also reproduced by the potential.
$E_d$   and  the
phase shifts do  not  depend  on  the  parameters  of  interaction
$V^{s}_{00}$ and $V^{t}_{00}$. Nevertheless, the triton binding energy $E_t$
depends on $V^{s}_{00}$ and $V^{t}_{00}$.
Thus, varying $V^{s}_{00}$ and $V^{t}_{00}$  we  managed  to  fit  the
potential simultaneously to the most important two- and three-body
observables ($E_d$, $<r^{2}_{0}>^{1/2}$, singlet and triplet $S$-wave $NN$
scattering
phases and $E_t$) without making use of three-body forces.
%     REFERENCES

\end{document}